\RequirePackage{fix-cm}
\documentclass[twocolumn,epjc3]{svjour3}  
\smartqed  
\RequirePackage{mathptmx}
\RequirePackage{graphicx}
\RequirePackage{subfigure}
\RequirePackage{rotating}
\RequirePackage{pdflscape}
\RequirePackage{color}
\RequirePackage{lineno}
\RequirePackage{latexsym}
\RequirePackage{xspace}
\RequirePackage[numbers,sort&compress]{natbib}
\RequirePackage[colorlinks,citecolor=blue,urlcolor=blue,linkcolor=blue]{hyperref}
\usepackage{amsmath,amsfonts,bm}
\usepackage{wasysym}
\usepackage{upgreek}
\usepackage{rotating}
\begin{document}

\title{Disentangling the sources of ionizing radiation in superconducting qubits}

\author{
L.~Cardani\thanksref{INFN-Roma}\and
I.~Colantoni\thanksref{INFN-Roma, CNR}\and
A.~Cruciani\thanksref{INFN-Roma}\and
F.~De~Dominicis\thanksref{GSSI,LNGS}\and
G.~D'Imperio\thanksref{INFN-Roma}\and
M.~Laubenstein\thanksref{LNGS}\and
A.~Mariani\thanksref{INFN-Roma}\and
L.~Pagnanini\thanksref{GSSI,LNGS,QUEEN}\and
S.~Pirro\thanksref{LNGS}\and
C.~Tomei\thanksref{INFN-Roma}\and
N.~Casali\thanksref{INFN-Roma}\and
F.~Ferroni\thanksref{INFN-Roma,GSSI}\and
D.~Frolov\thanksref{FNAL}\and
L.~Gironi\thanksref{Milano,INFN-Milano}\and
A.~Grassellino\thanksref{FNAL}\and
M.~Junker\thanksref{LNGS}\and
C.~Kopas\thanksref{Rigetti}\and
E.~Lachman\thanksref{Rigetti}\and
C.~R.~H.~McRae\thanksref{NIST,Boulder,Colorado}\and
J.~Mutus\thanksref{Rigetti}\and
M.~Nastasi\thanksref{Milano,INFN-Milano}\and
D.~P.~Pappas\thanksref{Rigetti}\and
R.~Pilipenko\thanksref{FNAL}\and
M.~Sisti\thanksref{INFN-Milano}\and
V.~Pettinacci\thanksref{INFN-Roma}\and
A.~Romanenko\thanksref{FNAL}\and
D.~Van~Zanten\thanksref{FNAL}\and
M.~Vignati\thanksref{INFN-Roma,Sapienza}\and
J.~D.~Withrow\thanksref{UFL}\and
N.~Z.~Zhelev\thanksref{Northwestern}
}

\thankstext{e1}{e-mail: ambra.mariani@roma1.infn.it }
\thankstext{e2}{e-mail: francesco.dedominicis@gssi.it }

\institute{INFN, Sezione di Roma, P.le Aldo Moro 2, I-00185, Rome, Italy \label{INFN-Roma} \and
Consiglio Nazionale delle Ricerche, Istituto di Nanotecnologia, c/o Dip. Fisica, Sapienza Università di Roma, 00185, Rome, Italy \label{CNR} \and
Gran Sasso Science Institute, 67100, L'Aquila - Italy \label{GSSI} \and
INFN, Laboratori Nazionali del Gran Sasso, I-67100 Assergi (AQ), Italy \label{LNGS} \and 
Department of Physics and Engineering Physics Astronomy, Queen’s University Kingston, Ontario, K7L 3N6 Kingston, Canada \label{QUEEN}
\and
Fermi National Accelerator Laboratory, Batavia, IL 60510, USA \label{FNAL}
\and
Dipartimento di Fisica, Universit\`{a} di Milano-Bicocca, I-20126 Milano, Italy \label{Milano} 
\and
INFN, Sezione di Milano-Bicocca, I-20126 Milano, Italy \label{INFN-Milano} 
\and
Rigetti Computing, 2919 Seventh Street, Berkeley, CA 94710, USA \label{Rigetti} 
\and
National Institute of Standards and Technology, Boulder, Colorado 80305, USA \label{NIST} \and
Department of Physics, University of Colorado, Boulder, Colorado 80309, USA \label{Boulder} \and
Department of Electrical, Computer and Energy Engineering, University of Colorado Boulder, Boulder CO 80309 \label{Colorado} \and
Dipartimento di Fisica, Sapienza Universit\`a di Roma, P.le Aldo Moro 2, I-00185, Rome, Italy \label{Sapienza} \and 
Physics Department, University of Florida, US \label{UFL} 
\and
Center for Applied Physics and Superconducting Technologies, Northwestern University, Evanston, IL 60208 \label{Northwestern}
}

\date{Received: date / Accepted: date}

\maketitle

\begin{abstract}
Radioactivity was recently discovered as a source of decoherence and correlated errors for the real-world implementation of superconducting quantum processors. In this work, we measure levels of radioactivity present in a typical laboratory environment (from muons, neutrons, and $\gamma$-rays emitted by naturally occurring radioactive isotopes) and in the most commonly used materials for the assembly and operation of state-of-the-art superconducting qubits. We present a GEANT-4 based simulation to predict the rate of impacts and the amount of energy released in a qubit chip from each of the mentioned sources.  We finally propose mitigation strategies for the operation of next-generation qubits in a  radio-pure environment.
\end{abstract}



\section{Introduction}
\label{sec:intro}
The technology of macroscopic superconducting circuits offers many advantages in the development of quantum processors: ease in design, fabrication and operation, high fidelity, and fast gate-times. 
As proven by an increasing number of companies and research institutes, superconducting circuits are also primed to implement quantum entanglement, as they allow to inter-couple tens of qubits~\cite{RigettiAspen,Arute:2019,wu:2021,IBM:eagle}.
Both companies and research centers, are now aiming at a further scale-up in the number of entangled qubits. The work presented in this paper was done in the framework of the SQMS\footnote{https://sqms.fnal.gov} (Superconducting Quantum Materials and System) Center, an international effort towards high-performance quantum computing.

Cosmic rays, as well as the decay of naturally radioactive isotopes present in the laboratory and in sample materials, can interact with the qubit chip, releasing energy. Energy deposits produce electron/holes charges and, subsequently, phonons. Both, charges and phonons, diffuse in the chip with different footprints and interact with the qubits, resulting in a loss of coherence~\cite{Vepsalainen:2020,Wilen:2021} and, if multiple qubits are involved, in correlated errors~\cite{Wilen:2021,McEwen:2021}. Phonons in particular can be absorbed by the superconductor, breaking Cooper pairs into dissipative quasiparticles. It is well known that quasiparticles can poison superconducting qubits~\cite{Catelani_2011,Pop:2014,Vool_2014} (including emerging cat-qubits~\cite{Berdou:2022}) and superconducting microwave resonators~\cite{Patel_2017,deRooij2021}.
Moreover, new studies point to the existence of an interaction between ionizing radiation and the dominant qubit loss mechanism, namely two-level systems (TLSs)~\cite{Klimov:2018,deGraaf:2017,Burnett:2014,Bilmes:2020,Lisenfeld:2019,Martinis:2005,Muller_2019,McRae_2020}. According to Thorbeck et al.~\cite{Thorbeck:2022}, an energy deposit due to radioactivity causes frequency jumps in multiple TLSs, inducing fluctuations in the qubit lifetime and limiting the stability of the device.

Today, many groups are actively investigating strategies for the mitigation of the effects of radioactivity. Among the proposed ideas, we recall: (i) the suppression of radioactivity sources, benefiting from the experience of the scientific community developing low-radioactivity detectors for particle physics~\cite{Cardani:2021,Gusenkova:2022}; (ii) the development of ``traps" surrounding the qubit to protect it from travelling phonons~\cite{Henriques:2019,martinis:2021}; (iii) new processor design, in which matrices of qubits are deposited on chips decoupled from each other~\cite{gold:2021}; and (iv) assisting fault mitigation through the use of sensors located near physical qubits~\cite{PhysRevApplied.16.024025}. It is also worth mentioning emerging efforts in designing on-chip circulators to replace the present (bulky and full of ferrite) components and, at the same time, mitigate the tunneling of quasiparticles~\cite{PhysRevResearch.3.043211}. gcv

The work presented in this paper sets a first milestone for the SQMS Round Robin experiment, in which a standard multiqubit chip designed and fabricated by Rigetti Computing will be operated in multiple facilities belonging to the SQMS center (Boulder - Colorado, Fermilab and Northwestern University - Illinois, Rigetti - California, and the deep underground Gran Sasso Laboratories INFN-LNGS - Italy).
These measurements entail long time, interleaved T1 and T2 scans in order to distinguish between decoherence channels and obtain reproducible performance metrics~\cite{McRae:2021}, with the final goal of disentangling  all causes of decoherence and pinpointing new mitigation techniques.
The contribution of INFN-LNGS, in particular, is establishing a low-radioactivity environment to test the SQMS prototypes, starting with the Round Robin one.
In this framework, we foresee two main goals:
\begin{itemize}
    \item Estimating the suppression of the radioactivity level in the ``Round Robin" chip that can be obtained in the underground INFN-LNGS, in order to quantify the expected reduction before operating the prototype; 
    \item Establishing a path to ensure that the radioactivity level of INFN-LNGS will satisfy the requirements for the future SQMS prototypes.
\end{itemize}
For this purpose, we measured all known sources of radioactivity spanning from the environment (Section~\ref{sec:environment}) to the components used for qubit operation (Section~\ref{sec:materials}). We then used Monte Carlo simulations to model their effect on the Round Robin chip (Section~\ref{sec:simulations}). 

We highlight that, since radioactive interactions involve the qubit substrate (and not the qubit itself)~\cite{Cardani:2021,Wilen:2021}, the obtained results apply to most qubits, with only minor adjustments due to the substrate volume. 
The effect of the energy deposits in the substrate on qubit performance, on the contrary, depends on the implementation of the superconducting circuit.
The majority of superconducting qubits, indeed, is limited by TLS interactions. Radioactivity would become an issue only if this source of decoherence was mitigated. Nevertheless, some devices are already affected by the presence of quasiparticles or phonons
~\cite{Pop:2014,Grunhaupt2019,Gusenkova:2022,Berdou:2022}, such as those produced by ionizing interactions in the substrate. As a consequence, suppressing radioactivity would directly enhance the performance of these qubits.

\section{Setup of the prototype qubit}
\label{sec:setup}
The Round Robin chip is a 325\,$\mu$m--thick, 11.9$\times$7\,mm$^2$ high-resistivity silicon wafer hosting 16 transmon qubits (14 flux-tunable qubits, and two fixed-frequency qubits) with frequencies in the range 4.1 -- 4.9\,GHz, and readout resonators with  frequencies from 5.4 to 5.9\,GHz.
Since the vast majority of radioactive interactions involve the qubit substrate, we will only focus on the substrate and neglect the particular qubit design.

\begin{figure}[thb]
\begin{centering}
\includegraphics[width=0.9\columnwidth]{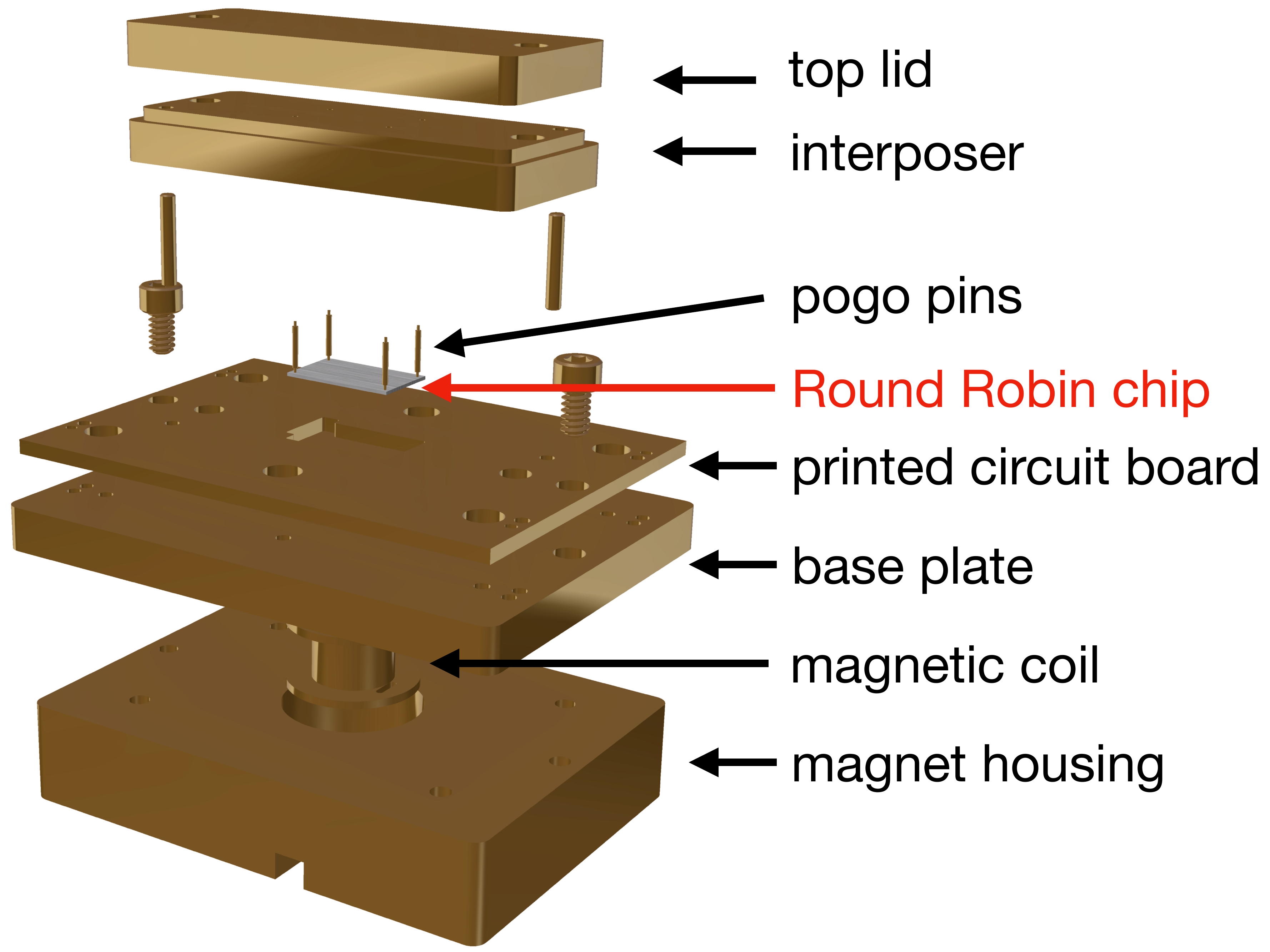}
\caption{Round Robin chip (gray) enclosed in its gold-plated copper box hosting the printed circuit board.}
\label{fig:setup1}
\end{centering}
\end{figure}

\begin{figure}[thb]
\begin{centering}
\includegraphics[width=\columnwidth]{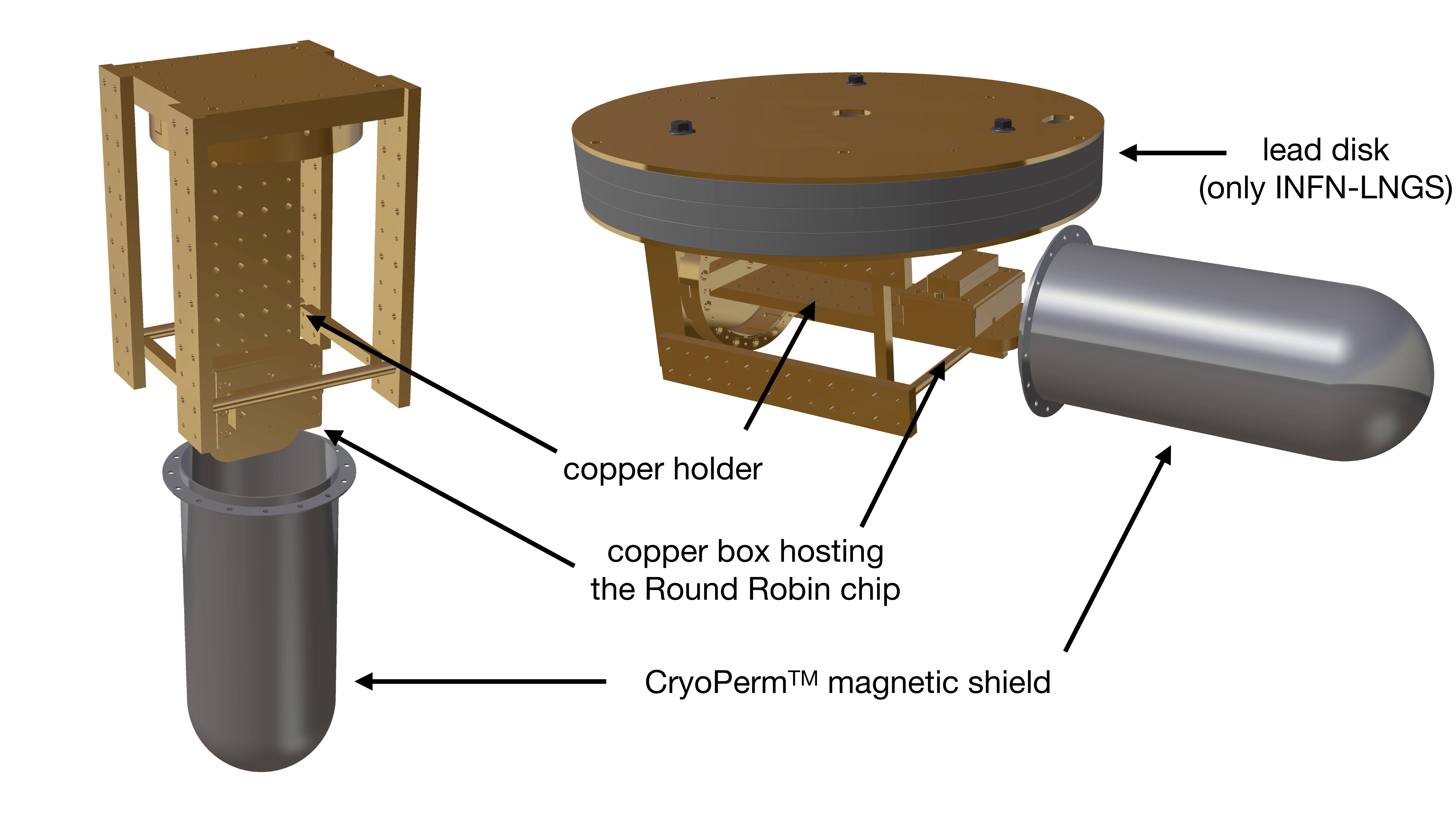}
\caption{Left: holder of the Round Robin chip. The copper box hosting the Round Robin chip, mounted on a gold-plated copper holder, is surrounded by a CryoPerm\textregistered\ magnetic shield. Gold-plated copper bars allow the installation of ``cold" electronics components (not shown in the picture) in the proximity of the sample. The square copper element can be mounted on the bottom of the mixing chamber plate (the coldest point of the dilution refrigerator). Right: installation in INFN-LNGS, where the DU experimental space prevents the vertical configuration. The picture shows also the lead shield that can be added in the INFN-LNGS facility.}
\label{fig:setup}
\end{centering}
\end{figure}

The chip is hosted in a copper box designed at NIST Boulder Laboratories, Colorado and its electrical connections are made using a printed circuit board (PCB). The box (Figure~\ref{fig:setup1}) is mounted on a gold-plated copper holder, which is thermally anchored to the coldest stage of the dilution refrigerator (DR) at $\sim$10\,mK, and enclosed in a CryoPerm\textregistered\ magnetic shield (Figure~\ref{fig:setup}). 

A variety of DRs will be used for these measurements at different institutes across the SQMS Center. Nevertheless, the materials used for the construction of the DUs and the dimensions of the vessels/components are rather similar, so the results we derived do not depend strongly on the particular design of the refrigerator. In the following simulations we modeled only the cryostat of INFN-LNGS.
The full experimental setup including the cryostat and the corresponding simulated geometry is shown in Figure~\ref{fig:simulated_setup}.

\begin{figure}[thb]
\begin{centering}
\includegraphics[width=\columnwidth]{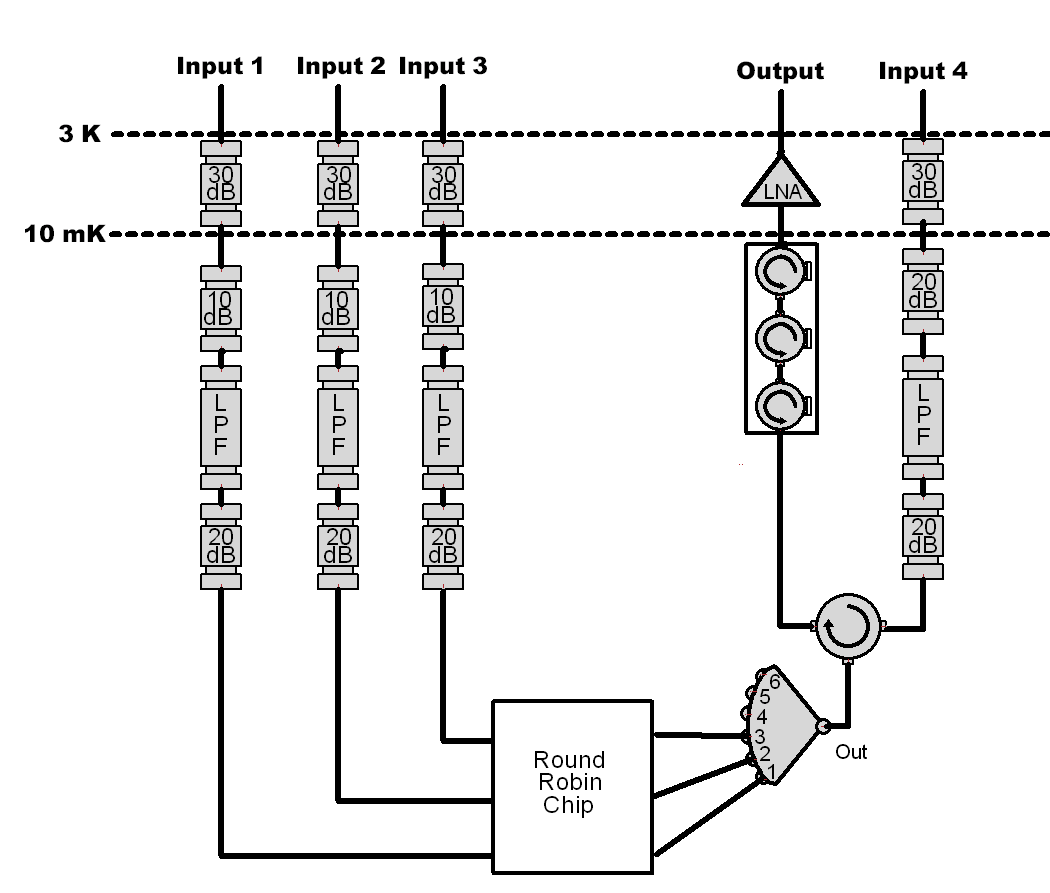}
\caption{Electronics setup for the Round Robin chip. The input lines are attenuated using 30\,dB attenuators at 3\,K and (10+20)\,dB attenuators at base temperature; moreover, they are filtered using 12\,GHz low-pass filters (``LPF"). The output signal is amplified using a 4 -- 8 GHz low noise cryogenic amplifier (``LNA"). A 6-channel cryogenic switch, low noise circulator and triple-junction isolator complete the readout.  }
\label{fig:readout_scheme}
\end{centering}
\end{figure}

Figure~\ref{fig:readout_scheme} shows a schematic view of the electronics components and readout to highlight their position compared to the qubit. 

Concerning the components that must be mounted close to the sample, we installed a Radiall\footnote{https://www.radiall.com} Cryogenic SP6T coaxial switch, a Low Noise Factory\footnote{https://www.lownoisefactory.com} triple junction isolator and a Low Noise Factory single junction circulator. For each input line, we installed two cryogenic XMA\footnote{https://www.xmacorp.com/} attenuators (10 dB  + 20 dB) and a 12\,GHz low-pass filter provided by K$\&$L Microwave\footnote{https://www.klmicrowave.com}. 
We anticipate that, in contrast to other facilities, the dilution refrigerator located in INFN-LNGS can be equipped with a lead shield at 10\,mK, placed between the sample and the electronic components (Figure~\ref{fig:setup}). 

To connect these components to the room-temperature electronics, we used coaxial cables made of copper beryllium (from 300\,K down to 3\,K), and NbTi (from 3\,K down to 10\,mK). The final connections from the mixing chamber plate to the sample are done using copper amagnetic coaxial cables. The input lines are attenuated at 3\,K using a 30\,dB XMA attenuator. The output signal is amplified using a Low Noise Factory 4 -- 8\,GHz cryogenic amplifier.


\section{Measurement of Far Radioactive Sources}
\label{sec:environment}
In the field of low-radioactivity detectors, it is common to distinguish between ``far" sources of radioactivity and ``close" sources of radioactivity, the main difference being that ``close" sources cannot be shielded as they are in the proximity of the device or part of the device itself.
As explained in this section, the INFN-LNGS offer a unique environment compared to other experimental sites involved in the Round Robin measurement, as the LNGS rock overburden naturally suppresses several ``far" sources of radioactivity by orders of magnitude. 

The most important sources of far radioactivity are cosmic rays, neutrons and $\gamma$-rays produced by radioactive decays in the laboratory environment. We underline that radioactive decays produce also other kind of particles, such as electrons and $\alpha$ particles. Nevertheless, the range of these particles in a medium-density material is of mm and $\mu$m respectively, and thus they cannot penetrate the cryostat vessels or the copper box in which the qubit is hosted. 

The main component of cosmic rays are muons. These ionizing particles arrive at sea level with an average energy of 4\,GeV and a typical flux of about 1\,$\mu$/cm$^2$/min~\cite{Pascale_1993}. 
The altitude of the experimental site, as well as the location of the laboratory within the site (e.g., if it is in the basement and thus shielded by the building itself) will impact the rate of muons reaching the device. 
The facilities involved in the Round Robin measurements are located about 0.05\,km (Rigetti), 0.20\,km (FNAL and Northwestern University) and 1.65\,km (Boulder - CU) above sea level. Considering that the cryogenic facilities of CU, FNAL and Rigetti are located at the first floor of buildings that are not heavily shielded, we expect the muon flux in FNAL (CU) to be a factor 1.1 (1.3) larger than at Rigetti.
The facility at Northwestern is being commissioned in the basement of the University, under five concrete floors. Thus, we expect a slightly lower muon flux compared to FNAL, despite the same elevation above sea level.\\
On the contrary, the cryogenic facility of INFN-LNGS is protected by a rock overburden of 1.4\,km, enabling a suppression of the muon flux by six orders of magnitude~\cite{MACRO:1995egd,Aglietta1998}. 

\begin{figure}[thb]
\begin{centering}
\includegraphics[width=\columnwidth]{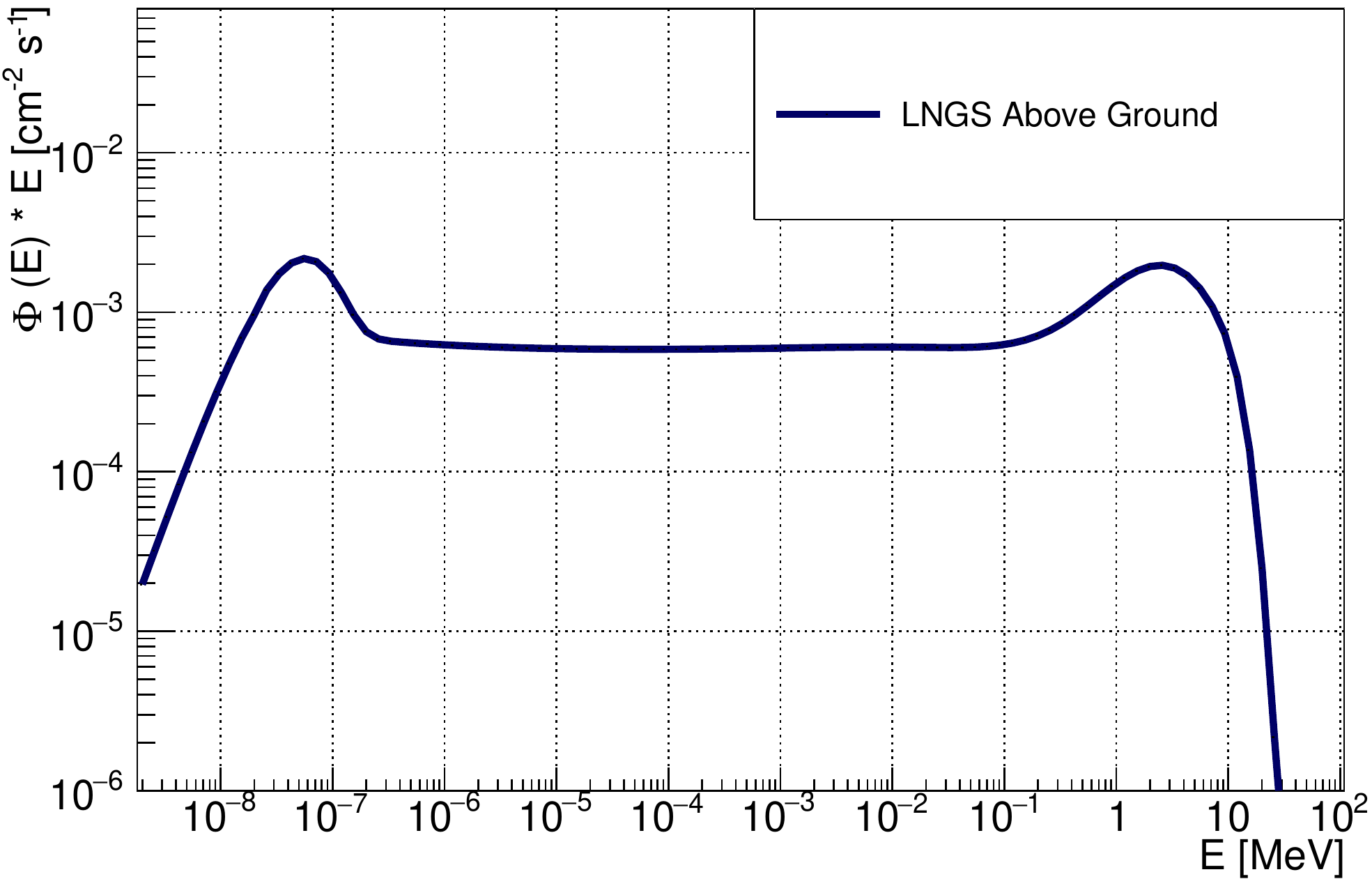}
\caption{Spectrum of neutrons measured above-ground at the INFN-LNGS. The total flux is 0.018\,n/cm$^{2}$/s, with an average experimental data uncertainty of 7\,$\%$.}
\label{fig:diamon}
\end{centering}
\end{figure}

The interaction of cosmic rays with our atmosphere also produces neutrons. Their flux at the surface extends from thermal energies (meV) to GeV, with an intensity that varies with altitude, geomagnetic field, and solar magnetic activity (\cite{Gordon} and references therein). In shallow and underground sites, low energy neutrons (below 10 MeV) are produced both by spontaneous fission and ($\alpha$,n) reactions, while fast neutrons are produced by nuclear reactions induced by residual cosmic ray muons in the rock or in the experimental apparatus.\\
We measured the flux of environmental neutrons in the energy region (0--20) MeV with a DIAMON neutron spectrometer \cite{POLA2020164078}, a portable detection system that provides real-time neutron spectrometry. 
The spectrum obtained \emph{above ground} at INFN-LNGS is shown in Figure~\ref{fig:diamon}. We measured a flux of 0.018\,n/cm$^{2}$/s, with an average experimental data uncertainty of 7\,$\%$.

The same neutron measurement was performed in other three laboratories located in different cities, obtaining negligible variations in the spectral shape, and a total flux spanning from 0.010 to 0.018\,n/cm$^{2}$/s\footnote{In a heavily shielded laboratory located at the Chooz Nuclear Power Station (France), we measured a neutron flux of 0.005\,n/cm$^{2}$/s. In a laboratory located in the Roma University - Sapienza basement, we measured 0.007\,n/cm$^{2}$/s. However, none of the laboratories involved in the Round Robin measurements is shielded by the building itself so we did not use these results.}. Thus, for the following study we will use the energy spectrum shown in Figure~\ref{fig:diamon} with an average flux of ($0.014\pm0.004$)\,n/cm$^{2}$/s, for all the sites located above ground.\\
On the contrary, the neutron flux in the underground INFN-LNGS has been measured to be several orders of magnitude lower than the atmospheric neutron flux~\cite{BEST20161}, on the order of 10$^{-6}$ n/cm$^{2}$/s, with the rate of muon-induced neutrons from two to three orders of magnitude lower than the rate of neutrons produced by nuclear reactions in the mountain rocks.

\begin{figure}[thb]
\begin{centering}
\includegraphics[width=\columnwidth]{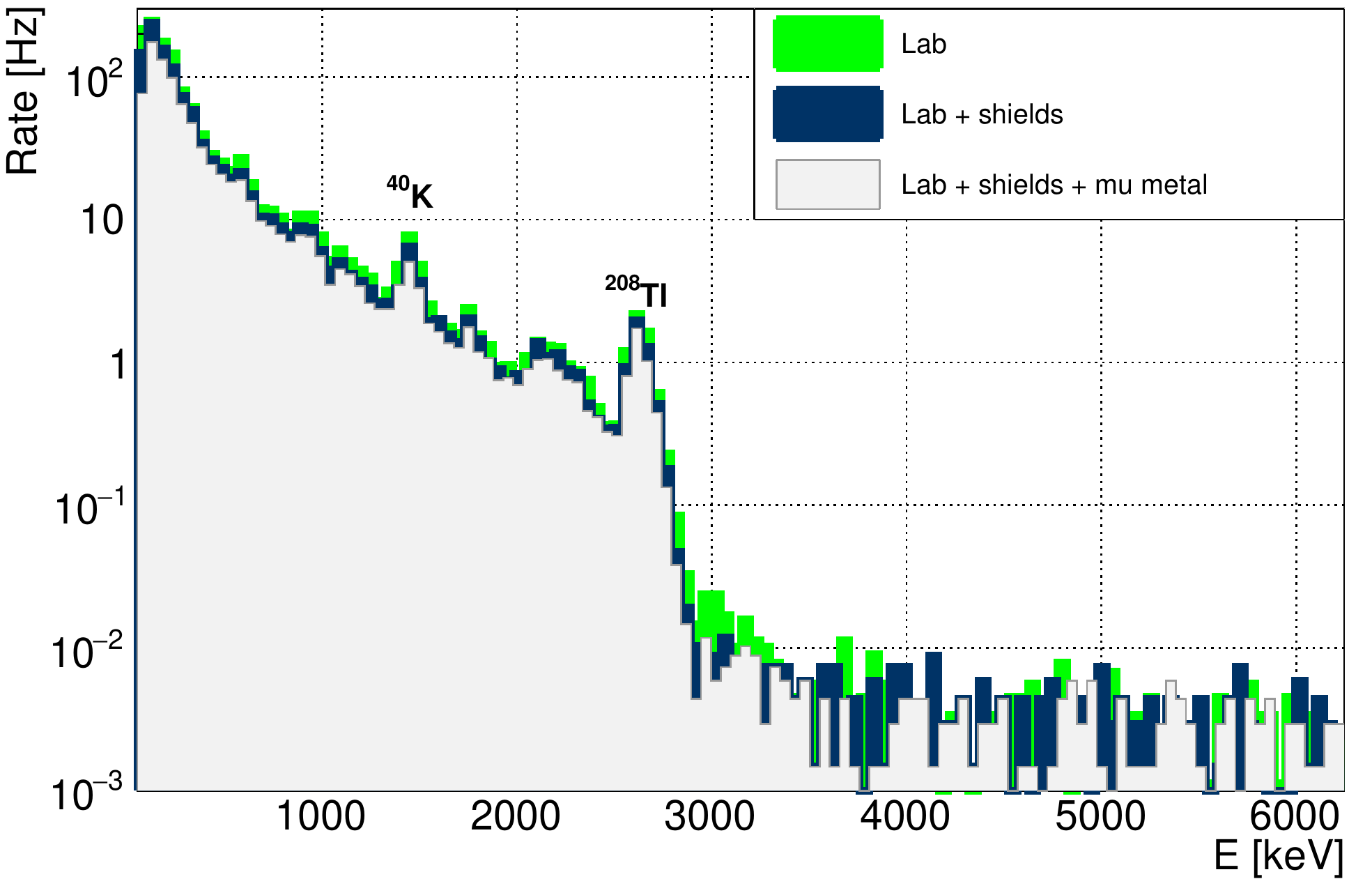}
\caption{Energy spectrum of the $\gamma$-ray flux measured using a 3" portable NaI spectrometer. Green: original spectrum collected in a laboratory of cryogenic detectors (Italy). Blue: same spectrum inside the cryostat thermal and vacuum vessels. Gray: same spectrum as (blue) but adding also the mu-metal magnetic shield.}
\label{fig:gammaNaI}
\end{centering}
\end{figure}

Environmental $\gamma$-rays have average fluxes of few $\gamma$/cm$^2$/sec and typical energies lower than 2.6\,MeV (one of the $\gamma$-rays produced by the decay of the $^{208}$Tl isotope).
Figure~\ref{fig:gammaNaI} shows the $\gamma$-rays spectrum measured with a 3" portable NaI spectrometer in a laboratory of cryogenic detectors (Roma, Italy).

As in the case of neutrons, the shape of the spectrum is expected to be more or less the same in the different Round Robin locations. Indeed, most of the radioactive isotopes naturally present in the environment belong to the $^{232}$Th- or $^{238}$U-chains (with the exception of the 1.4\,MeV peak due to the primordial radionuclide $^{40}$K).
On the other hand, the rate of each peak depends on the specific contamination of the laboratory, and on the total shielding. 
For this work, we measured the absolute flux in the experimental hall of INFN-LNGS and in other above ground laboratories, obtaining (1.0$\pm$0.5)\,$\gamma$/cm$^2$/sec and (2.5$\pm$0.5)\,$\gamma$/cm$^2$/sec respectively\footnote{This value is representative for a ``typical" laboratory environment. In buildings made of particulars materials (such as tuff rock) the $\gamma$-ray flux can be a factor of three higher.}.

Finally, Figure~\ref{fig:gammaNaI} shows three measurements performed in the same site to evaluate the effect of the DR itself on the chip. Typical DRs, indeed, include one or two vacuum cans, and thermal vessels to protect the samples from thermal radiation. In addition, they can be equipped with magnetic shields. 
We performed the same measurement in three different scenarios: (i) without vessels, (ii) closing the DR with its  thermal vessels and vacuum cans, and (iii) adding also a mu-metal magnetic shield. The $\gamma$-ray flux, obtained by integrating the gamma rate over the entire energy range, is reduced by a factor of 25$\%$ and 33$\%$ inside the vessels and inside the vessels + mu-metal respectively.
Our simulation (Section~\ref{sec:simulations}) accounts for such effects.

\section{Measurement of Close Radioactive Sources}\label{sec:materials}
Close sources of radioactivity comprise cables, electronic components, and other materials that cannot be placed far away from the chip.
In this work, we focus on the components that will be used for the Round Robin measurements. Nevertheless, the vast majority of these components are common in many experimental facilities and, thus, this information can be used to predict background contributions in different qubit prototypes. Our list includes:
\begin{enumerate}
    \item [\textit{A}] printed circuit board (PCB) - weight: $\sim$7\,grams. The PCBs used for the Round Robin were produced by San Francisco Circuits. They consisted of a 1.575$\times$2.205\,cm$^2$, 0.157\,cm thick 3-layer FR408HR (a high performance laminate for multi-layer Printed Wiring Board applications, where maximum thermal performance is required); 
    We also measured a new type of non-magnetic PCB, that however was not mounted in the current prototype of the Round Robin chip and is indicated with A* in Table~\ref{tab:Contaminations};
    \item[\textit{B}] gold-plated copper box (total mass: 0.4\,kg) and a gold-plated copper holder (2.2\,kg) used to thermally anchor the qubit to the mixing chamber plate as shown in Figure~\ref{fig:setup};
    \item[\textit{C}] ``cold" magnetic shield, consisting in a hollow CryoPerm\textregistered~cylindrical shield with 78.8\,mm diameter, 193.5\,mm height and 1.0\,mm thickness - weight: $\sim$475\,grams;
    \item [\textit{D}] Intelliconnect non-magnetic SMA adapters - weight: $\sim10$\,grams;
    \item [\textit{E}] Intelliconnect non-magnetic copper coaxial cables (diameter: 2.19 mm, length: 25 cm);
    \item [\textit{F}] Radiall  SP6T cryogenic switch - weight: $\sim$165\,grams;  
    \item [\textit{G}] 4-12\,GHz Low Noise Factory circulator (mod. CIC4\_12A), not used in the Round Robin measurements - weight: $\sim$50\,grams;   
    \item [\textit{H}] 4-12\,GHz Low Noise Factory dual-junction circulator (mod. CICIC4\_12A), not used in the Round Robin measurements - weight: $\sim$97\,grams;   
    \item [\textit{I}] 4-8\,GHz Low Noise Factory triple-junction isolator (mod. ISISISC4\_8A) - weight: $\sim$145\,grams;  
    \item [\textit{J}] 10-30\,dB XMA attenuators (mod. 2082-6418-10-CRYO, 2082-6418-20-CRYO, and 2082-6418-30-CRYO) - weight: $\sim$5\,grams each;   
    \item [\textit{K}] K$\&$L Low Pass Filters (mod. 3L250-12240/T20000 and 3L250-8160/T20000) - weight: $\sim$15\,grams each; 
    \item [\textit{L}] COAX
    \footnote{www.coax.co.jp} NbTi superconducting coaxial cables running from 4\,K to 15\,mK (mod SC-219/50-NbTi-NbTi). 

\end{enumerate}

Other components are placed at the 4\,K stage and thus, their contribution to the qubit counting rate is mitigated by their distance from the sample:
\begin{enumerate}
    \item [\textit{M}] 4 -- 8\,GHz Low Noise Factory cryogenic amplifier - weight: $\sim$17\,grams;  
    \item [\textit{N}] COAX copper-beryllium coaxial cables running from 300\,K to 4\,K (mod SC-119/50-B-B).
\end{enumerate}

\begin{table}
\centering
\caption{Bulk contamination - values and 90\% C.L. upper limits - of the materials located in the proximity of the qubit at the 15 mK stage (A -- L) and 4\,K (M,N) stage. Items indicated with (*) were measured for general interest but will not be used in the Round Robin measurements. The full results are reported in the Supplemental Material and summarized in this table: for the $^{232}$Th decay chain, we quote the maximum activity between the ones measured for $^{228}$Ra and $^{228}$Th; concerning $^{238}$U and daughters, we quote the $^{226}$Ra activity, being representative of the most worrisome part of the decay chain. }
\begin{center}
\resizebox{.5\textwidth}{!}{
\begin{tabular}{cccccc}
\hline\noalign{\smallskip}
Component   & $^{232}$Th          & $^{238}$U         & $^{235}$U       & $^{40}$K           & $^{137}$Cs  \\
            & [mBq/kg]            & [mBq/kg]          & [mBq/kg]        & [mBq/kg]           & [mBq/kg] \\
\hline
\textit{A}  &  $(18000 \pm 1000)$ & $(11500 \pm 400)$ & $(710 \pm 110)$ & $(12000 \pm 1000)$ & $<30$ \\
\textit{A*} &  $(5410 \pm 330)$   & $(4200 \pm 200)$  & $(230 \pm 50)$  & $(4200 \pm 500)$   & $<40$ \\
\textit{B}  &  $<1.5$             & $<1.2$            &  $<4$           &  $<9$              &  $<0.6$ \\ 
\textit{C}  &  $<8.4$             & $<8.3$            &  $<8.4$         &  $<35$             &  $<2.7$ \\ 
\textit{D}  &  $(46 \pm 13)$      & $(42 \pm 10)$     &  $(70 \pm 30)$  &  $(240 \pm 90)$    &  $<10$ \\ 
\textit{E}  &  $(54 \pm 12)$      & $(44 \pm 11)$     &  $(34 \pm 17)$  &  $(740 \pm 130)$   &  $<12$ \\ 
\textit{F}  &  $(1880 \pm 100)$   & $(1340 \pm 60)$   & $(130 \pm 30)$  &  $ (2200 \pm 300)$ &  $<11.2$ \\
\textit{G}  &  $<310$             & $<330$            &  $<410$         &  $<2000$           &  $<60$ \\ 
\textit{H*} &  $<250$             & $<380$            &  $<380$         &  $<2600$           &  $<60$ \\ 
\textit{I*} &  $<190$             & $<240$            &  $<220$         &  $<2000$           &  $<50$ \\ 
\textit{J}  &  $<52$              & $(200 \pm 20)$    &  $<47$          &  $<140$            &  $<13$ \\ 
\textit{K}  &  $(23 \pm 4)$       & $<9.1$            & $(60 \pm 10)$   & $<100$             & $<1.9$ \\
\textit{L}  &  $<750$             & $<1000$           &  $<380$         &  $<7000$           &  $<230$ \\ 
\hline
\textit{M}  &  $<890$             & $<1000$           &  $<850$         &  $<10000$          &  $<210$ \\
\textit{N}  &  $(240 \pm 40)$     & $<78$             &  $(350 \pm 90)$ &  $<500$            &  $<20$ \\
\hline
\textit{O*} & $(53 \pm 4)$      & $(9400 \pm 900)$&  $(350 \pm 30)$   &  $(290 \pm 40)$ &  $< 2.2$ \\
\textit{P*} & $<10$   & $<11$   &  $<4.5$   &  $<87$ &  $<5$ \\
\hline
\end{tabular}
\label{tab:Contaminations}}
\end{center}
\end{table}

Finally, we report the results for a thermally conductive epoxy glue (Stycast\textregistered, O*) and DOW CORNING cryogenic grease (P*). These materials are widely used for qubit application and their radioactive content is of general interest. However, we did not make use of them for the Round Robin chip. 

We investigated the radiopurity of these components using the $\gamma$-spectrometric techniques detailed in the Supplemental Materials.
The results of the screening, reported in Tab.~\ref{tab:Contaminations}, show that the PCBs (A) contain a significant amount of natural radioactivity. This was expected, due to the PCBs composition (i.e. glass fiber)~\cite{Aprile:2019,Armengaud_2017,ITO2020163050}. 
We also include a possible alternative type of a-magnetic PCB (A*), consisting of copper (three layers, for a total thickness of 87\,$\mu$m) interleaved by dielectric layers (hydrocarbon ceramic laminates) produced by ROGERS Corporation\footnote{https://www.rogerscorp.com} under the codes 4350B (two layers, for a total thickness of 420\,$\mu$m) and 4450F (a single layer, 95\,$\mu$m thick).
As shown in Table~\ref{tab:Contaminations}, these PCBs (A*) were about a factor of 3 cleaner, from the radioactivity point of view, compared to those selected for the Round Robin chip. 

On the contrary, the materials dominant in weight and closest to the chip - copper (B) and CryoPerm (C) - feature a good radiopurity level. Nevertheless, these components are the ones most subject to cosmogenic activation. Indeed, this analysis confirms the presence of cosmogenic radioisotopes - $^{54}$Mn, $^{57}$Co, $^{58}$Co, and $^{60}$Co - produced above-ground by cosmic ray interaction with the materials~\cite{Baudis:2015kqa,Cebrian:2017oft}. The results are reported in Tab.~\ref{tab:Cosmo}. 
\begin{table}
\centering
\caption{Activity concentration - values and 90\% C.L. upper limits - for short- and long-lived radioisotopes produced by cosmogenic activation in the materials of the detector setup.} 
\begin{center}
\resizebox{.5\textwidth}{!}{
\begin{tabular}{ccccc}
\hline\noalign{\smallskip}
Component & $^{60}$Co  & $^{58}$Co & $^{57}$Co & $^{54}$Mn\\
          & [mBq/kg]     & [mBq/kg]   & [mBq/kg]   & [mBq/kg]\\
\hline
\textit{B}    &  $(0.6 \pm 0.3)$   & $(4 \pm 1)$   &  $-$   &  $(2.4 \pm 0.8)$\\ 
\textit{C}    &  $<3.7$   & $(14 \pm 5)$  &  $(20 \pm 7)$ &  $ - $ \\ 
\textit{D}    &  $(51 \pm 8)$  & $ - $   &  $ - $   & $ - $ \\
\textit{K}    &  $(5 \pm 1)$  & $ - $   &  $ - $   & $ - $ \\
\hline
\end{tabular}
\label{tab:Cosmo}}
\end{center}
\end{table}

\section{Effects on the Qubit Chip}
\label{sec:simulations}
\begin{figure*}[thb]
\begin{centering}
\includegraphics[width=1.5\columnwidth]{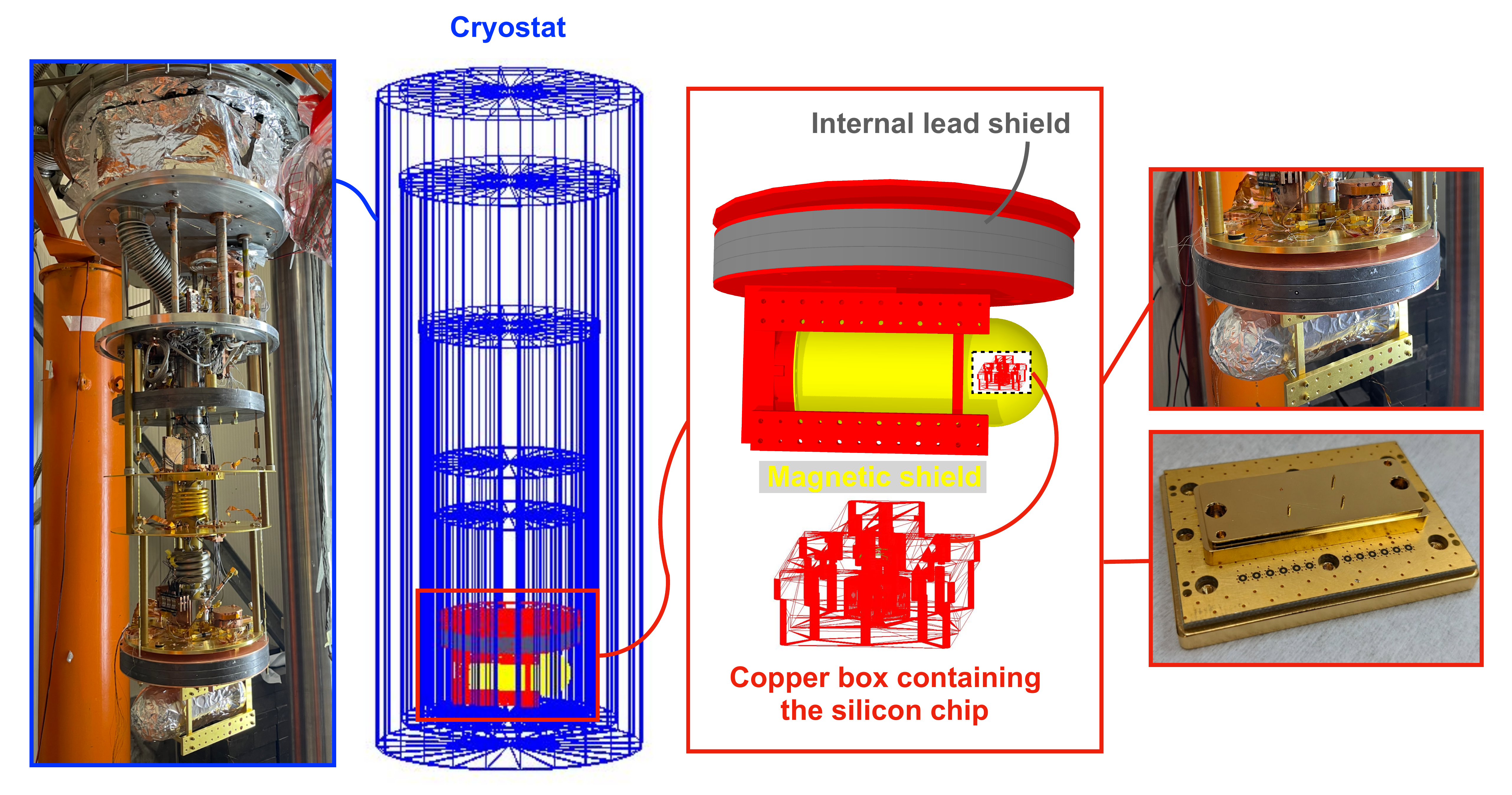}
\caption{Experimental setup as implemented in the simulation. The chip inside the copper box is the only active volume. The copper box, holder and magnetic shield were imported from the CAD file, while for the dilution refrigerator we implemented a simplified version.}
\label{fig:simulated_setup}
\end{centering}
\end{figure*}

The geometry of the setup described in Figures~\ref{fig:setup1} and \ref{fig:setup} was implemented in our GEANT-4 based simulation framework~\cite{AGOSTINELLI2003250}. We also implemented the copper holder, the magnetic shield, and a simplified version of the dilution refrigerator, including its internal and external lead shields (Figure~\ref{fig:simulated_setup}).
The Round Robin chip was considered the only active volume, where we stored tracks and recorded the energy deposited by each simulated interaction.
In the Monte Carlo simulations two different approaches were chosen for far and close sources.

For the far sources ($\gamma$-rays, muons and neutrons), we generated primary particles within the laboratory environment.
More in detail, for neutrons and $\gamma$-rays we generated the events uniformly distributed on the surface of a cylinder enclosing the cryostat, according to the measured energy spectra (Figures~\ref{fig:diamon} and \ref{fig:gammaNaI}) and with isotropic momentum distribution.
Concerning muons, in order to reproduce the actual angular distribution, we randomly generate positions on an hemisphere around the cryostat according to a cos$^2$ distribution\footnote{In INFN-LNGS the angular distribution is affected by the rock overburden and cannot be accurately described by the cos$^2$ distribution. This does not impact significantly the result.}. For every sampled position, we generate muons perpendicularly from a (120$\times$120)\,cm$^2$ tangent-plane to the hemisphere.

We then estimate the rate of interactions in the chip for each of these sources by scaling the number of recorded events to the flux measurements (Section \ref{sec:environment} and references therein), both for the above-ground laboratories (``standard") and the underground INFN-LNGS. \\
Simulations were done considering the three possible INFN-LNGS shielding setups: no shield at all, only the external shield surrounding the cryostat (10\,cm thick wall of lead bricks), and the so-called ``full" shield configuration, with both the external lead shield and the inner lead shield (3\,cm thick lead disk between the chip and the mixing chamber).
The results are summarized in Table~\ref{tab:ResultsFarCotamination}.
\begin{table}[!ht]
    \centering
       \caption{Interaction rate in the substrate of the Round Robin chip. For the above-ground facilities we assumed a ``standard" $\gamma$-ray flux of (2.5$\pm$0.5)\,$\gamma$/cm$^2$/sec and a muon flux of 1\,$\mu$/cm$^2$/min. For the LNGS facility we used the measured flux of (1.0$\pm$0.5)\,$\gamma$/cm$^2$/sec and we evaluated the suppression factor due to the presence of lead shields.}
    \begin{tabular}{lcccc}
    \hline\noalign{\smallskip}
     Source        &``standard"   &LNGS  &LNGS  &LNGS\\
                   &               &     &Ext. Shield &Full Shield\\
                   &$\big[$mHz$\big]$ & $\big[$mHz$\big]$ & $\big[$mHz$\big]$ & $\big[$mHz$\big]$\\
    \hline
    Lab $\gamma$-rays &(18$\pm$4)       &(7.0$\pm$3.5)  &(0.10$\pm$0.05) &(0.07$\pm$0.04)\\
    Muons          &(10.0$\pm$0.6)       &$<$10$^{-5}$  &$<$10$^{-5}$ &$<$10$^{-5}$\\
    Neutrons       &(0.15$\pm$0.05)  &$<$10$^{-4}$  &$<$10$^{-4}$ &$<$10$^{-4}$\\
    \end{tabular}
    \label{tab:ResultsFarCotamination}
\end{table}

Concerning above-ground facilities, the highest interaction rate in the Round Robin chip is produced by $\gamma$-rays from laboratory radioactivity, resulting in a rate of (18$\pm$4)\,mHz. 
The rate in INFN-LNGS is only (7.0$\pm$3.5)\,mHz, due to intrinsically lower content of natural radioactivity in the Gran Sasso rock. This rate is further suppressed through the combination of the external and internal lead shields. More in detail, the external shield alone and the internal shield alone  abate the $\gamma$-ray flux by a factor 70 and 2 respectively. Their combination offers a flux suppression by a factor 100, for a final value of  (0.07$\pm$0.04)\,mHz.
This suppression factor is based on the assumption that the external lead shield is an ideal cylinder 80\,cm tall, 10\,cm thick. Less tight geometries could limit the suppression capability of the external shield.
The intrinsic radioactivity of the lead shields could in principle generate interactions in the Round Robin chip. To check this, we simulated a typical contamination of $^{210}$Pb with an activity concentration of 100\,Bq/kg in the internal lead shield (the most worrisome one, due to its proximity with the sample) and obtained a negligible interaction rate in the Round Robin chip of 0.01\,mHz.\\
In conclusion, this shielding system is very effective in suppressing the dominant radioactive source and could be easily installed also in above ground facilities\footnote{The shielding capability in above-ground facilities would be slightly less effective due to the presence of high energy $\gamma$-rays produced by cosmic rays.}.

The second most significant contribution to the rate of the Round Robin chip ((10.0$\pm$0.6)\,mHz in the Rigetti laboratories) is due to muons. 
Despite their lower rate, muons are considered more worrisome than $\gamma$-rays, as they produce long tracks potentially affecting more qubits lying on the same chip~\cite{Wilen:2021}.

The only viable strategy to suppress muons is moving the facility to an underground site. As shown in Table~\ref{tab:ResultsFarCotamination}, INFN-LNGS offers an extremely high reduction of cosmic rays; nevertheless, effective suppression factors can be obtained even in shallow sites. As an instance, the shallow underground laboratory NEXUS (Northwestern Experimental Underground Site at Fermilab, Illinois, U.S.) is shielded by only 100\,m of soil/gravel, and yet offers a suppression by three orders of magnitude of atmospheric muons~\cite{Ren_2021,battaglieri:2017}. 

An alternative strategy may consist of tagging muons (instead of suppressing them) in order to identify and reject operations done while these particles are traversing the chip. The drawback of this approach is that a compact and efficient external muon veto to be operated above-ground would feature a trigger rate as high as hundreds of Hz~\cite{Wagner_2022}. More sophisticated vetoes at cryogenic temperature could guarantee a similar efficiency while diminishing the trigger rate (and thus, the associated dead time).

We also recall that other groups are proposing to suppress the devastating  effects of muons at the chip level by proposing novel distributed error correction schemes~\cite{Xu.2022}.\\

Finally, we observe that neutrons, that are the main concern for classical computers~\cite{Baumann:2005}, have a negligible impact rate in typical superconducting qubits.\\

“Close” sources of radioactivity were simulated by generating the radioactive decays of the relevant isotopes uniformly distributed within the volume of each component (A-N). The location of every component in the simulated geometry was assumed as the typical place where the given item is usually mounted in the cryostat. Due to the proximity of some components to the Round Robin chip (PCB and copper box in particular), we simulated also $\beta$- and $\alpha$-particles, in addition to $\gamma$-rays.
We then estimated the rate of interactions in the chip from each component by scaling the number of recorded events to the measured activity concentrations reported in Table~\ref{tab:contamination_details} in the Supplemental Material (and summarized in Table~\ref{tab:Contaminations}).

\begin{table}
    \centering
 \begin{tabular}{lcc}
    \hline\noalign{\smallskip}
    Component  & Description        & Rate   \\
               &                    &$\big[$mHz$\big]$   \\
    \hline

    \textit{A} & PCB                &4.52 $\pm$ 0.04   \\
    \textit{B} & Box                & [1 -- 6]$\times$10$^{-3}$   \\
    \textit{B*}& Holder             & [2 -- 4]$\times$10$^{-4}$   \\
    \textit{C} & Magnetic Shield    & [2 -- 9]$\times$10$^{-4}$   \\
    \textit{D} & {\bf SMA}             & $(2 \pm 0.4)\times 10^{-5}$   \\
    \textit{E} & {\bf Cu coax cables}  & $ (3 \pm 0.6) \times 10 ^{-5}$  \\
    \textit{F} & Cryogenic switch   & $(1.0 \pm 0.2) \times 10 ^{-2}$ \\
    \textit{G}& Circulator         & $<8\times10^{-4}$  \\
    \textit{H*}& Dual-junct. circulator   & $<2\times10^{-3}$   \\
    \textit{I*} & Triple-junct. isolator  & $<2\times10^{-3}$   \\
    \textit{J} & {\bf Attenuators}             & [0.5 -- 1]$\times$10$^{-5}$  \\
    \textit{K} & {\bf Low Pass Filters}        & $(1 \pm 0.2) \times 10^{-5}$   \\
    \textit{L} & {\bf NbTi cables}             &  $<4 \times 10^{-4}$ \\
    \textit{M} & Cryogenic amplifier      & {$<2\times10^{-5}$}   \\
    \textit{N} & {\bf Cu-Be cables}       & {1$\times$10$^{-6}$ }    \\
    \hline
    \end{tabular}
    \caption{Interaction rate in the substrate of the Round Robin chip. The copper box and holder were split in this Table to highlight that, due to its proximity to the qubit, the contribution of the box is more important than the one of the holder, despite the same level of radiopurity. Some components are highlighted in bold to remind that usually these components are placed in large quantity in the dilution refrigerator (and the rate should be scaled accordingly).}
    \label{tab:MaterialiSimulazione}
\end{table}

The results, summarized in Table~\ref{tab:MaterialiSimulazione}, show that the vast majority of electronics components are not worrisome for qubit measurements, with the exception of the PCB. Being very close to the qubit and, at the same time, not very radiopure, the PCB constitutes an irreducible source of interactions in the Round Robin chip. 
This problem is well known to particle physicists developing detectors for rare events searches, as both fiberglass and ceramic (typical materials used for the PCB multilayers) contain usually levels of radioactivity comparable to those in rocks and soils. The presence of the PCB is the ultimate limit to the improvement of the radioactivity level in the INFN-LNGS laboratory environment, compared to above-ground sites. 

\begin{figure}[thb]
\begin{centering}
\includegraphics[width=\columnwidth]{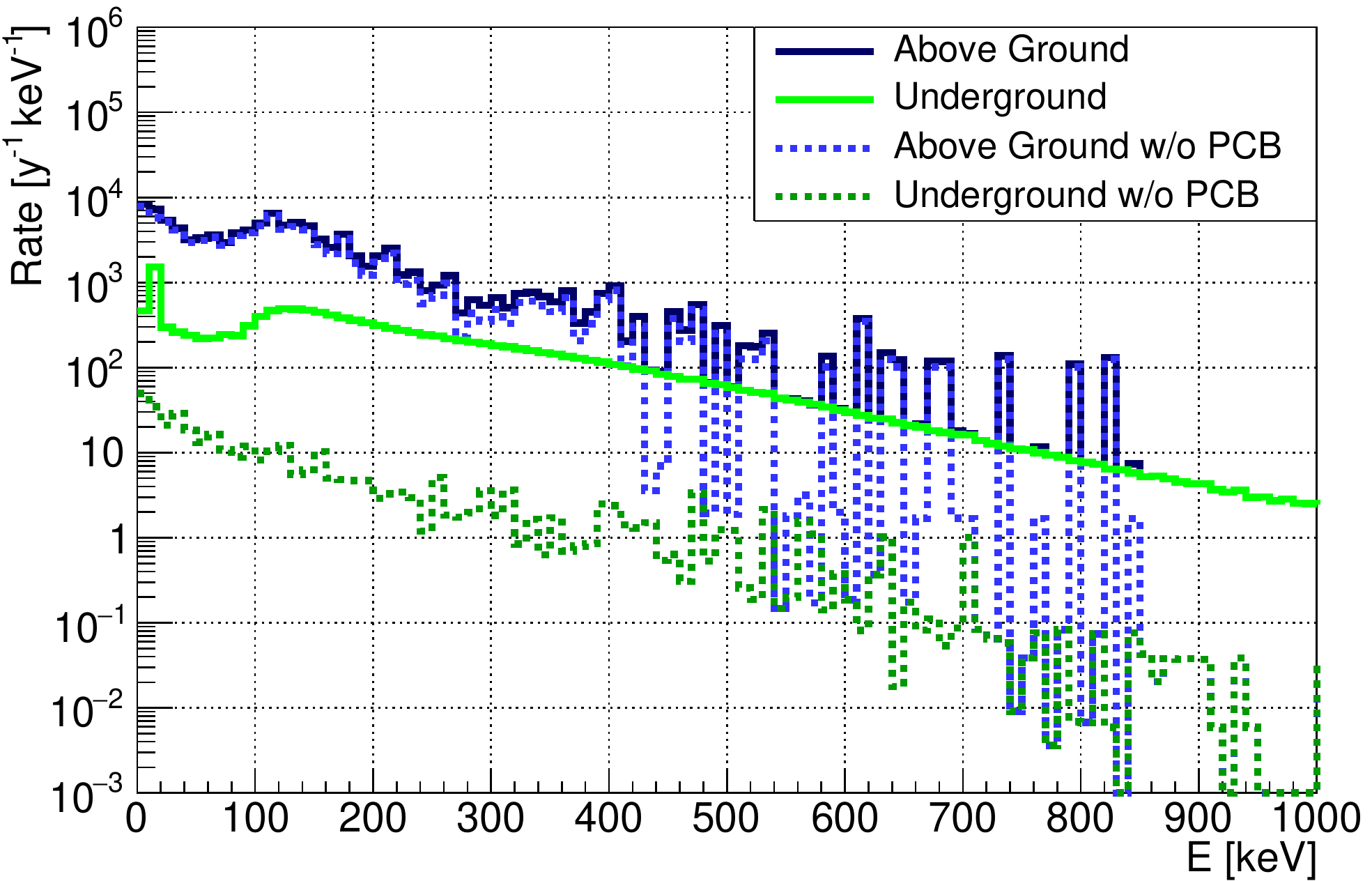}
\caption{The rate of interactions in the Round Robin chip is reported as a function of the energy that they deposit in the silicon chip. The majority of interactions deposit tens/hundreds of keV. Solid lines: total rate expected in a typical above ground laboratory (blue) and in the INFN-LNGS underground site (green). Dashed lines show what would be the rate in the two cases without the PCB contribution. Above ground, the effect of the PCB is almost negligible, as the rate is dominated by far radioactive sources. On the contrary, in the deep underground INFN-LNGS the PCB is the major contribution to the interaction rate. As a consequence, substituting the PCB with a more radio-pure one would allow to abate the rate of interactions in the chip.}
\label{fig:results_simu}
\end{centering}
\end{figure}

In Figure~\ref{fig:results_simu}, we report the rate of events as a function of the energy that they deposit in the Round Robin chip. 
In this plot we distinguish the results expected for the above-ground laboratories and for INFN-LNGS, where the overall rate is expected to be about an order of magnitude lower.
In both cases we show the results with and without the PCB (the only relevant contribution of the ``close" sources), to show the improvement that might be obtained by substituting this component with a more radio-pure one.

We want to emphasize that our screening measurements (Section~\ref{sec:materials}) are not sensitive to fast decaying isotopes, i.e. radioisotopes that are activated above-ground and decay within seconds or minutes. In principle, these radioisotopes could further increase the interaction rate in the chip when the Round Robin is operated above-ground. Nevertheless, we simulate the effect of the cosmogenic activation of copper (the most ``massive" material in the setup) using the ACTIVIA software~\cite{BACK2008286} and we obtained a negligible rate in the chip of 0.05\,mHz. Furthermore, some of the components (G, H, I, M) were measured also above ground, finding no evidence of activated isotopes within the reported sensitivities. 

This study proves that the radioactivity suppression obtained in INFN-LNGS can be considered satisfactory for the Round Robin measurements, as it allows us to search for macroscopic differences in the prototype performance.

For a long-term goal of reaching qubits lifetime of a seconds (the goal of the 3-D SQMS architecture), it is likely that further improvements will be required. Indeed, with the measured contamination levels in INFN-LNGS, the probability of observing a radioactivity-induced event in a second-long time window, amounts to 0.4$\%$. Nevertheless, such a rate was estimated using the geometry of the Round Robin chip. A chip hosting 256 qubits will likely demand a larger substrate and thus be prone to a higher rate of interactions, requiring lower radioactivity levels. The PCB in particular could become the main issue in suppressing the interaction rate.
As explained in Section~\ref{sec:materials}, we already measured the content of natural radioactivity of a new type of non-magnetic PCB, obtaining a factor $\sim$3 improvement compared to the Round Robin PCBs (Table~\ref{tab:Contaminations}). A further improvement, if needed, will require a dedicated R$\&$D activity.

\section{Conclusions}
In this work, we measured the radioactive content of all the components that are commonly used in the characterization of superconducting qubits. We considered a particular case (the Round Robin chip of the SQMS center) to run Monte Carlo simulations predicting the impact of each radioactive source on the qubit chip.

We conclude that the overall interaction rate is dominated by $\gamma$-rays radioactivity of the laboratory environment, that can be suppressed using lead shields.

The second contribution comes from muons. Even if muon-vetoes, or on-chip mitigation strategies can be envisioned, the most effective abatement strategy for these interactions consists in operating the prototypes in underground sites, such as the INFN-LNGS for the SQMS center.

Finally, the setup components resulted generally radio-pure, with the exception of PCBs, that demand a dedicated optimization. Our studies already identified an intermediate solution, allowing to suppress the interaction rate coming from the PCBs by roughly a factor of 3, but other solutions may be necessary for next-generation quantum processors.

\begin{acknowledgements}
The authors thank the Director and technical staff of the Laboratori Nazionali del Gran Sasso. We are grateful to A.~Girardi for the controller of the cryogenic switch, M.~Guetti for the help with the whole cryogenic facility (including its continuous upgrades), and M.~Iannone for the design and construction of the chip holder and readout.
We are also grateful to the LNGS Computing and Network Service for computing resources and support on U-LITE cluster at LNGS.
We thank F. Cappella and G. Pessina for the stimulating discussions on the simulator of muons and on the radioactivity of the PCBs, respectively.
The work was supported by the U.S. Department of Energy, Office of Science, National Quantum Information Science Research Centers, Superconducting Quantum Materials and Systems Center (SQMS) under the contract No. DE-AC02-07CH11359, and by the Italian Ministry of Research under the PRIN Contract No. 2020h5l338 (“Thin films and radioactivity mitigation to enhance superconducting quantum processors and low temperature particle detectors”). 
Finally, we are grateful to Raffele Tripiccione (Lele) for his tireless effort in building this collaboration.
\end{acknowledgements}

\bibliographystyle{spphys}       

\section*{SUPPLEMENTAL MATERIAL}

\subsection{Technical implementation of the $\gamma$-rays simulation}
The simulation of ``far" radioactive sources is complicated due to the poor statistics obtained with the simulations. The probability for a $\gamma$-ray produced in the laboratory to reach a chip as small as a $\sim$cm$^2$, is indeed extremely small (we expect about one interaction every 10$^8$ simulated events).

\begin{figure}[thb]
\begin{centering}
\includegraphics[width=0.5\columnwidth]{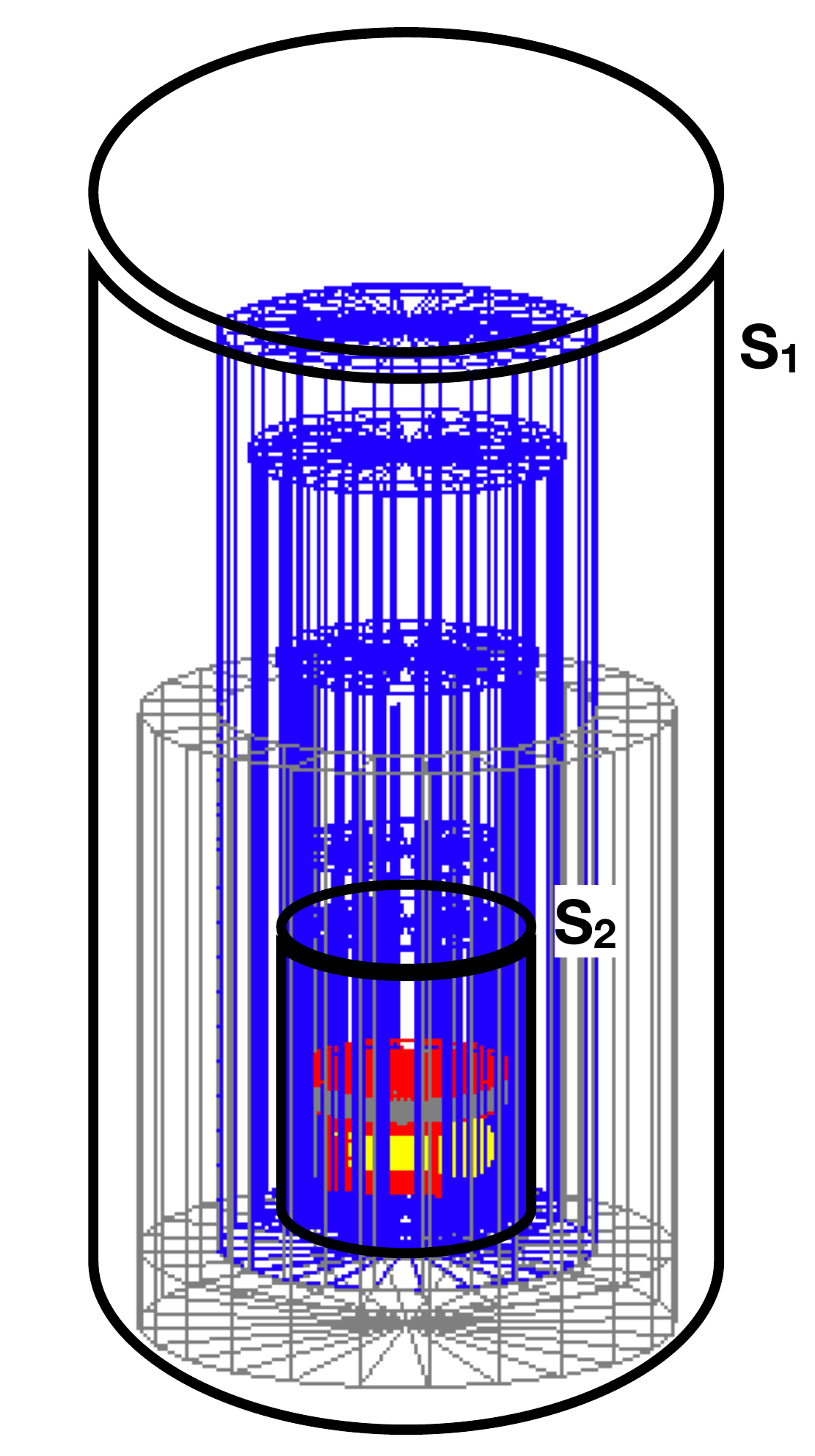}
\caption{Experimental setup as implemented in the ``full" shield configuration, with both an internal 3-cm thick lead disk and an external 10-cm thick lead shield. The two surfaces used to generate $\gamma$-rays in the double-step simulation are shown.}
\label{fig:2step_sim}
\end{centering}
\end{figure}

To increase the statistics we implemented a double-step simulation.
First, we generated 10$^7$ $\gamma$-rays from a cylindrical surface (S$_1$, Fig.~\ref{fig:2step_sim}) enclosing the cryostat and located just outside the external lead shield. Such gammas were generated according to a $\gamma$-ray spectrum measured with a 3" portable NaI spectrometer in a laboratory of cryogenic detectors (Italy) and with isotropic momentum distribution. We registered the kinetic information (E, p$_x$, p$_y$, p$_z$) for gamma-rays entering a second cylindrical surface (S$_2$, Fig.~\ref{fig:2step_sim}), that encompasses the mixing chamber of the cryostat. 

\begin{figure}[thb]
\begin{centering}
\includegraphics[width=\columnwidth]{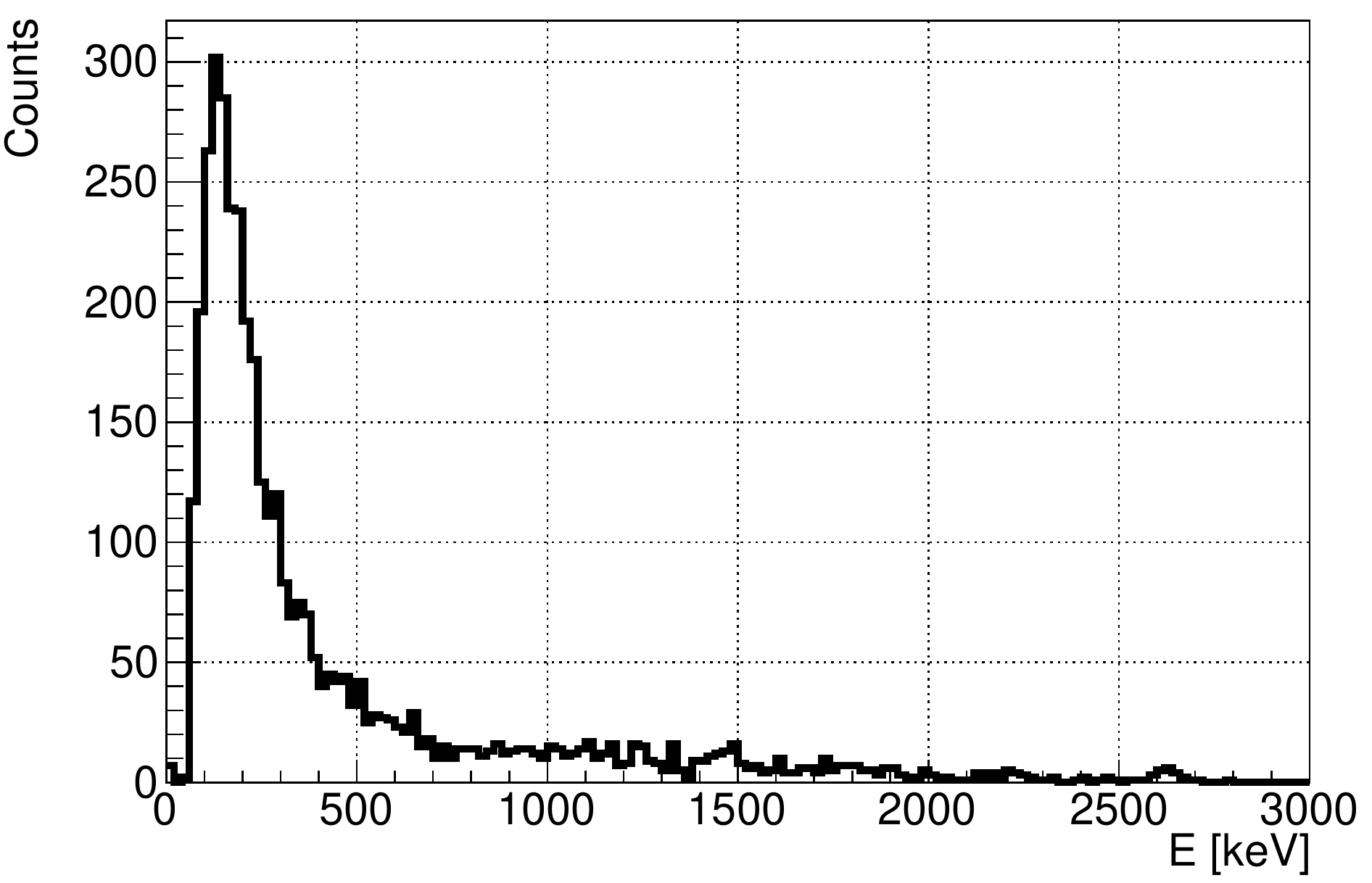}
\includegraphics[width=\columnwidth]{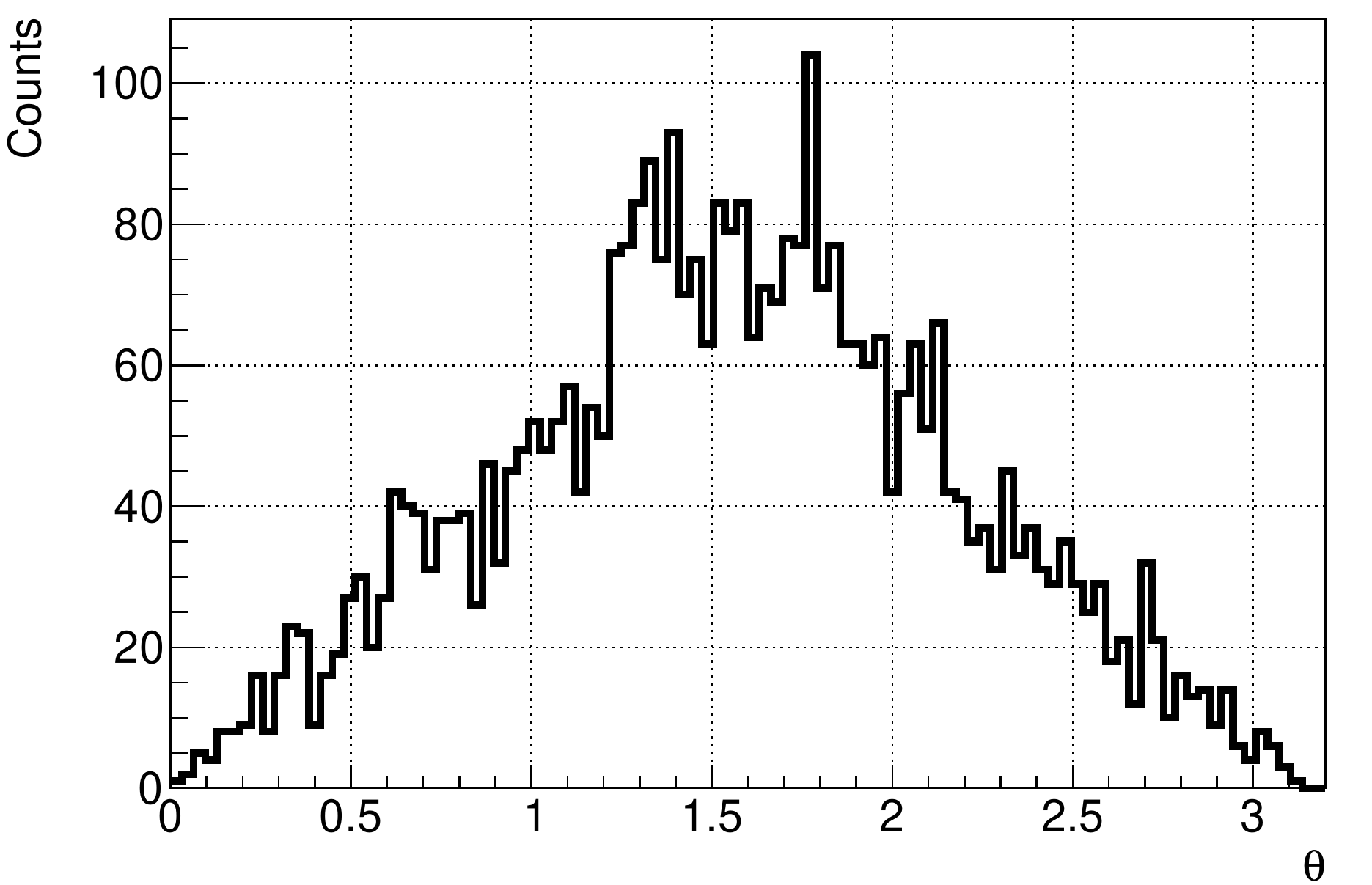}
\caption{Energy spectrum (top) and polar angular distribution (bottom) recorded for gammas entering S$_2$.}
\label{fig:S2_RecordedDistributions}
\end{centering}
\end{figure}

In a second step, we generated a few 10$^8$ gammas from S$_2$ according to the energy spectrum and the polar angular distribution ($\theta$) recorded in the first step of the simulation, as shown in Fig.~\ref{fig:S2_RecordedDistributions} for the ``full" shield case. 

Finally, to obtain the rate of impacts in the chip, we divided the number of events depositing energy in the chip by the equivalent time of the simulation, which is given by t$_\mathrm{eq}$ = N$_\mathrm{gen}$/(A$_2$ $\cdot$ flux$_2$), where N$_\mathrm{gen}$ is the number of generated events, A$_2$ is the area of the cylindrical surface used to generate gammas in the second step, and flux$_2$ is the gamma-ray flux calculated in S$_2$ after the first step.

\subsection{Detailed $\gamma$-spectrometry results}
In this section, we report a more detailed list of results of the radio-assay of the ``close" materials. The samples were measured with High-Purity Germanium (HPGe) detectors installed underground in the STELLA (SubTerranean Low-Level Assay) facility at INFN-LNGS, and above-ground at the Radioactivity Laboratory of Milano - Bicocca University.

The HPGe detectors used at INFN-LNGS are coaxial p-type germanium detectors with an active volume of about 200 - 400 cm$^2$ and an optimized design for high counting efficiency in a wide energy range. The energy resolution of the spectrometers is about 2.0 keV at the 1332 keV line of $^{60}$Co. To reduce external background, the detectors are shielded with a $\sim$20 cm layer of low-radioactivity lead, copper ($\sim$5 cm) and a 5 cm layer of Polyethylene. Each set-up is continuously flushed with high-purity boil-off nitrogen to prevent radon entering the shield. The HPGe detector used at Milano-Bicocca University is a coaxial p-type germanium in low-background configuration, with a 30\% relative efficiency (about 200 cm$^2$ active volume) and an energy resolution of 1.7 keV (Full Width at Half Maximum) at the 1332 keV line of $^{60}$Co. It is surrounded by 5 cm of copper plus 5 cm of lead to reduce the external background. We placed each sample directly above the end-cap of the HPGe detector to maximize the detection efficiency. We exploit a Monte Carlo tool based on GEANT4 to simulate the various experimental configurations and estimate the detector efficiency with a precision between 5\% and 10\%, as explained in Ref.~\cite{BORIODITIGLIOLE2014249}. More details on experimental set-ups, detector performance, results, and Monte Carlo validation of the detector performances can be found in Refs.~\cite{Laubenstein:2020rbe,Bellini:2020alj,Beretta:2021msd}. 

For each $\gamma$-ray emitting isotope belonging to the natural radioactive decay chains, we quote the value and the 68\% C.L. uncertainty of its activity. In the latter, we include the statistical uncertainty of both peaking signal and subtracted background, and a conservative 5-10\% uncertainty on the efficiency. When multiple $\gamma$s-lines are emitted by isotopes in secular equilibrium, we statistically combine the activity of all the involved lines. In cases in which no evidence of a peak was found, we placed a 90\% upper limit on the activity. More details about the adopted analysis procedure is reported in Ref.~\cite{HEISEL2009741}. 

For each detector component, we simulated all isotopes separately and scaled the obtained rate of interactions by the activities reported in Table~\ref{tab:contamination_details}. The sum of all the scaled rates was reported in Table~\ref{tab:MaterialiSimulazione}.

\begin{table*}[]
    \centering
    \caption{Full results - values and 90\% C.L. upper limits - 
    of the material screening performed at LNGS with Ultra-Low Background High Purity Germanium detectors. We investigate the radiopurity of the elements located in the proximity of the qubit at the 15 mK stage and 4\,K stage. Moreover, items G, H, O and P were measured but have not been used in the Round Robin measurements. All the activities are reported in mBq/kg.}
    \label{tab:contamination_details}
    \resizebox{\textwidth}{!}{
    \begin{tabular}{ccccccccccc}
    \hline
    Chain & Isotope &  PCB(A) & Copper (B) & CryoPerm\textregistered~(C) & SMA adapters (D) & Cu coax. cable (E) & Cryo-switch (F) & Circulator (G) & Circulator (H) \\
    \hline
    $^{232}$Th	& $^{228}$Ra &	$18000 \pm 1000$ &	$<0.8$ &	$<4.7$ &	$<48$ &	$51 \pm 15$ &	$1750 \pm 110$ &	$<190$ &	$<240$ \\
	& $^{228}$Th &	$18000 \pm 1000$ &	$<1.5$ &	$<8.4$ &	$46 \pm 13$ &	$54 \pm 12$ &	$1880 \pm 100$ &	$<310$ &	$<250$ \\
    $^{232}$U	& $^{234}$Th &	$11000 \pm 2000$ &	$<160$ &	$<500$ &	$1800 \pm 600$ &	$1500 \pm 400$ &	$4600 \pm 600$ &	- & 	-\\
	& $^{234m}$Pa&	$10000 \pm 2000$ &	$<25$ &	$<1400$ &	$1700 \pm 600$ &	$1000 \pm 400$ &	$3300 \pm 700$ &	- &	- \\
	& $^{226}$Ra &	$11500 \pm 400$ &	$<1.2$ &	$<8.3$ &	$42 \pm 10$ &	$44 \pm 11$ &	$1340 \pm 60$ &	$<330$ &	$<380$ \\
	& $^{235}$U &	$710 \pm 110$ &	$<4.1$ &	$<8.4$ &	$70 \pm 30$ &	$34 \pm 17$ &	$130 \pm 30$ &	$<410$ &	$<380$ \\
	& $^{40}$K&	$12000 \pm 1000$ &	$<9$ &	$<35$ &	$240 \pm 90$ &	$740 \pm 130$ &	$2200 \pm 300$ &	$<2000$ &	$<2600$ \\
	& $^{137}$Cs &	$<30$ &	$<0.6$ &	$<2.7$ &	$<10$ &	$<12$ &	$<11.2$ &	$<60$ &	$<60$ \\
\hline
Chain	& Isotope &	Isolator (I)& Attenuators (J) &	Low pass filters (K) &	NbTi coax. cable (L) & Amplifier (M) &	CuBe coax. cable (N) &	Stycast\textregistered~(O) & Vacuum Grease (P)\\
\hline
$^{232}$Th	& $^{228}$Ra &	$<190$ &	$<52$ &	$<18$ & $<800$ &	$<890$ &	$210 \pm 50$ &	$52.8 \pm 4.4$ & $<10$\\
	& $^{228}$Th &	$<190$ &	$38 \pm 11$ &	$23 \pm 4$ & $<750$ &	$<880$ &	$240 \pm 40$ &	$52.7 \pm 4.2$ & $<6.3$	\\
$^{232}$U	& $^{234}$Th &	- &	$<1100$ &	$1700 \pm 200$ &-&-&	$6000 \pm 1000$ & $9400 \pm 900$ & $<100$ \\
	 & $^{234m}$Pa&	- &	$<2100$ &	$1100 \pm 200$ &-&-&	$8000 \pm 3000$ & $8100 \pm 800$ &	$<210$\\
	& $^{226}$Ra &	$<240$ &	$200 \pm 20$ &	$<9.1$ & $<6000$&	$<1000$ &	$<78$ &	$47.5 \pm 2.8$ &	$<11$\\
	& $^{235}$U &	$<220$ &	$<47$ &	$60 \pm 10$ &$<380$&	$<850$ &	$350 \pm 90$ & $350 \pm 30$ & $<4.5$\\
	& $^{40}$K&	$<2000$ &	$<140$ &	$<100$ &$<7000$&	$<10000$ &	$<500$ & $290 \pm 40$ &	$<87$\\
	& $^{137}$Cs &	$<50$ &	$<13$ &	$<1.9$ &$<230$&	$<210$ &	$<20$ &	$<2.2$ & $<5.0$\\
    \hline
    \end{tabular}}
\end{table*}

\end{document}